# Broadband Terahertz Frequency Comb Based on Actively Mode Locked Resonant Tunneling Diode

Feifan Han[1,4], Hongxin Zhou[3,4], Qun Zhang[1], Zebin Huang[1], Longhao Zou[1], Weichao Li[1], Fan Jiang[1], Jingpu Duan[1], Jianer Zhou[1], Xiongbin Yu[1,(a)], Zhen Gao[3,(b)], and Xiaofeng Tao[2,1,(c)]

[1]Department of Broadband Communication, Pengcheng Laboratory, Shenzhen, China
[2]National Engineering Research Center of Mobile Network Technologies, Beijing University of Posts and Telecommunications, Beijing, China
[3]Department of Electronic and Electrical Engineering, Southern University of Science and Technology, Shenzhen, China
[4]These authors contributed equally: Feifan Han, Hongxin Zhou

(a)Corresponding author: yuxb@pcl.ac.cn
(b)Corresponding author: gaoz@sustech.edu.cn
(c)Corresponding author: taoxf@bupt.edu.cn

**Abstract**

The frequency combs characterized by their phase-coherent equidistant spectral modes and precise frequency scales of broadband spectrum, have made them an indispensable part of contemporary physics. A terahertz (THz) frequency comb is a key asset for THz technology applications in spectroscopy, metrology, communications, and sensing. However, the THz frequency comb technologies are comparatively underdeveloped compared to the optical frequency domain, primarily attributed to the deficiency of advanced THz generation components. In this paper, we innovatively demonstrate a compact THz frequency comb source based on a resonant tunneling diode (RTD) through active mode locking technique. By injecting a strong continuous-wave radio frequency (RF) signal via the bias line into a RTD oscillator integrated within a WR-5 hollow metallic waveguide package, we observe a broadband comb spectrum spanning from 140 to 325 GHz. The mode spacing is directly determined by the frequency of the injected RF signal, providing a wide tuning range of approximately 40 GHz. We also employ the proposed frequency comb source as the local oscillator in a coherent transmitter. In particular, this is the first all-electrical compact THz frequency comb source, and the transmission demonstration paves the way to next-generation communication.

## I. Introduction

Frequency combs, characterized by a series of evenly spaced, coherent spectral lines in the frequency domain, have played a pivotal role in advancing terahertz (THz) sources. By down-conversion of optical frequency comb to the THz range, the performance of THz sources has been significantly enhanced, achieving unprecedented levels of phase noise reduction, spectral purity and frequency stability[1]. However, such systems typically rely on bulky optical synthesis setups. In contrast, direct generation of THz frequency combs using semiconductor devices offers a compact and integrated alternative. In recent years, quantum cascade lasers (QCLs) have made remarkable progress in THz frequency combs generation, utilizing both passive and active mode-locking techniques[2-3]. Nevertheless, THz QCLs generally require cryogenic cooling due to low lasing efficiency. Despite recent advancements in room-temperature THz frequency combs by QCL using difference-frequency generation method, frequency combs below 1 THz remains challenging[4]. Alternatively, resonant tunnelling diodes (RTDs) operating in the frequency range of 0.1 to 2 THz at room temperature[5], have shown promise for frequency combs generation via passive mode locking[6]. However, the inherent limitations associated with passive mode locking, such as cumbersome optical setups and sensitivity to external feedback, pose challenges for realizing compact and tunable room-temperature THz frequency comb sources. In this work, we present the first demonstration of RTD-based THz frequency comb generation through active mode locking. By injecting a strong radio frequency (RF) signal into the RTD, broadband frequency comb generation is achieved without the need for external optical feedback. Moreover, the mode spacing of the comb can be widely tuned by varying the RF injection frequency. The frequency comb performance is evaluated in terms of frequency stability and phase noise characteristics. The measured phase noise of beat note frequency is below -100 dBc/Hz at 100 Hz offset. Ultimately, a coherent transmitter based on the proposed RTD-based frequency comb source is further implemented for multi-channel data transmission. A maximum data rate of 12 Gbps per spectral channel is successfully demonstrated through 16-quadrature amplitude modulation (16-QAM) using online digital signal processing.

**II. Design**

Figure 1 illustrates the conceptual design of the THz frequency comb source based on an RTD oscillator packaged into WR-5 hollow metallic waveguide. RTD oscillator consists of an active region, a split ring resonator (SRR) and a shunt resistor terminated slot resonator. The active region is composed of a sandwiched AlAs/InGaAs/AlAs heterostructure grown on Indium phosphide (InP) substrate, forming a dual-barrier, single-quantum-well structure configuration that enables the resonant tunneling effect. This effect introduces a negative differential conductance (NDC) in the current–voltage (*I-V*) characteristics of the RTD, resulting in strong nonlinearity and significant gain in the THz frequency range. The SRR and the slot resonator function together as an E-plane coupler, facilitating the coupling of THz signals between the RTD and the WR-5 waveguide. To generate frequency comb, a strong RF signal is injected into the RTD via the bias line, driving the device into an active mode-locking regime. Owing to the intrinsic nonlinearity and gain of RTD at THz frequency region, multiple modes with evenly spaced frequencies are generated, thereby forming a broadband THz frequency comb. Each comb mode frequency $f_n$ is an integer multiple of the injected RF frequency $f_{RF}$, expressed as $f_n = n \times f_{RF}$.

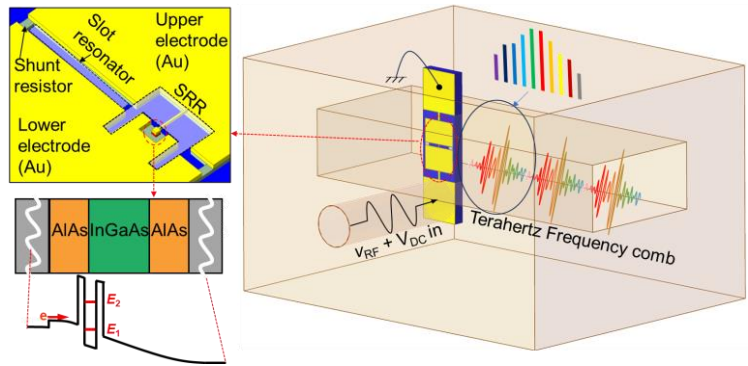

Fig.1 THz comb source with an RTD oscillator chip is integrated into WR-5 hollow metallic waveguide.

### III. Operation Mechanism

The RTD oscillator consists of an *RLC* resonator, implemented using a THz antenna, and a RTD functioning as a gain and nonlinear element, as illustrated in Figure 2a. The resonant tunneling effect in RTD gives rise to a distinctive feature in the *I-V* characteristic of the RTD, and provide the nonlinear and gain mechanism, shown in Fig. 2b. Specifically, the *I-V* curve exhibits NDC, indicated by the negative slope within the gray region. In this NDC region, an increase in voltage leads to a decrease in current. Consequently, when the bias voltage is set within this region, the RTD functions effectively as a gain and nonlinear element. When RF signal with amplitude *M* represented as $V(t)=M\cdot\cos(\omega t)$ is injected to the *RLC* resonator, THz frequency comb occurs.

To simplify the analysis, we approximate the *I-V* characteristics of the RTD using a third-order polynomial[5]:

$$I_{RTD}(V) = -\alpha V + \beta V^3$$

in which $\alpha, \beta > 0$, and the bias point is at the origin of the voltage and current axes.

Taking the time derivative of the Kirchhoff's current function, we obtain the following equation:

$$C\cdot\frac{d^2V}{dt^2} + \{-(\alpha - G_{loss} - G_{rad}) + 3\beta V^2\}\frac{dV}{dt} + \frac{V}{L} = \frac{d(M\cdot\cos(\omega t))}{dt}$$

This process highlights the complexity of the internal dynamics of the system and its potential oscillatory modes. After transforming this equation into a group of first order differential equations, we can obtain the voltage $V(t)$ across the RTD over time via backward differentiation methods[7]. Given our primary interest in the frequency domain, the mode distributions can be further analyzed through the Fast Fourier Transform function.

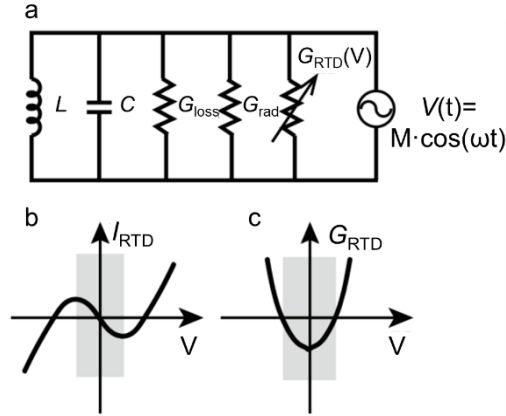

Fig. 2. RTD oscillator modeled as *RLC* resonator with RF injection. (a), Circuit diagram. (b), *I-V* curve of RTD approximated by third-order polynomial. (c), Differential conductance of RTD in the model. The gray area represents the region with NDC.

Fig. 3 illustrates the corresponding mode distributions with varied *M* from 0 to 1000. Starting without any RF signal input (M=0), Fig. 3(i) observes a typical emission power spectrum of a continuous-wave oscillatory state, characterized by a single-frequency spectrum with significantly lower power spectral densities in the sidebands compared to the main peak. When a weak RF signal is injected into RTD, two first-order sidebands around the main peak are typically generated in Fig.

3(ii). As stronger RF signals are applied, the RTD oscillator is actively mode locked, multiple modes with evenly spaced frequencies are produced due to increased energy availability from the external source. From Fig. 3(iii) to Fig. 3(v), the comb spectrum features wider frequency spanning and more uniform power distribution, benefiting from more energy transferred to the high frequency region.

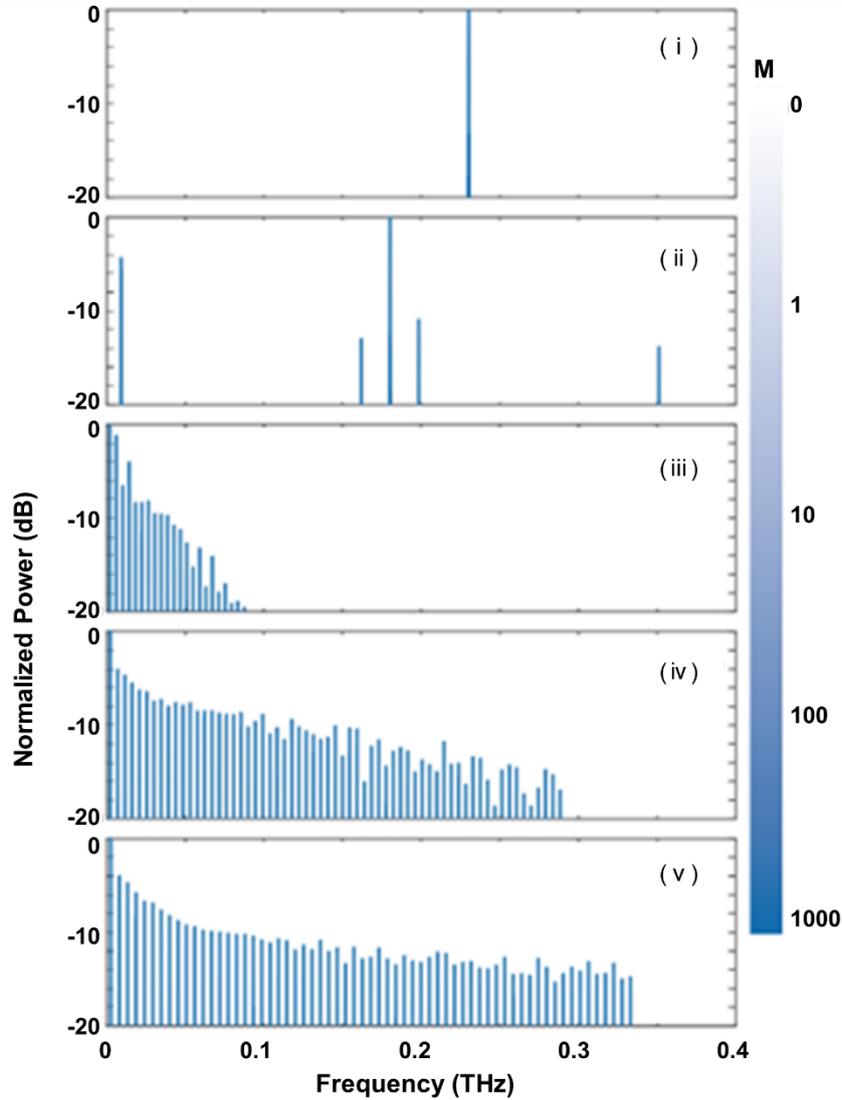

Fig. 3. Frequency comb distributions with a fixed RF frequency of ω=0.05, but varying amplitude of RF indices from 0 to 1000.

**IV. Results**

The fabricated THz comb source, integrating an InP-based RTD oscillator into a WR-5 waveguide, is shown in Fig. 4. The WR-5 waveguide is CNC-milled from two copper blocks forming the upper and lower parts. A 2.92 connector is inserted into upper part, and provides a direct connection to RTD oscillator with wire bonding. The 100 μm thickness RTD oscillator is positioned at the end of the metallic waveguide. The RTD epitaxial structure and fabrication process follow the method described in Ref.8.

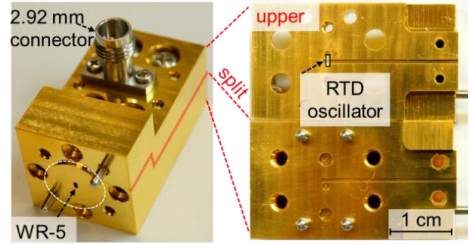

Fig.4 The fabricated THz comb source with RTD oscillator packaged with WR-5 waveguide.

We down converted the THz frequency comb and fed it into the spectrum analyzer for measurement. We used two frequency extenders working at 140-220 and 220-325 GHz respectively to obtain broadband measurement. Fig. 5a shows the measured spectrum of frequency comb spanning from 140 to 325 GHz, obtained with RF injection power of 18 dBm and $f_{RF}$ = 0.918, 2, 3, and 4 GHz respectively. The spectral lines between the comb modes are noise during measurement and can be suppressed when RBW of spectrum analyzer is reduced. We also observed the frequency comb extending beyond 40 GHz, which is the upper limit of the signal generator. The mode spacing is precisely equal to $f_{RF}$ and can be arbitrary controlled by varying the RF frequency, ranging from 0.8 GHz to over 40 GHz. The spectral envelope is primarily shaped by the frequency characteristics of the RTD oscillator and the WR-5 waveguide. Within the 140–190 GHz range, the comb modes exhibit nearly uniform power, resulting in a relatively flat spectrum. In contrast, the comb modes between 190 and 325 GHz show significantly reduced power due to the cutoff behavior of the WR-5 waveguide with the RTD oscillator. The magnified spectra at 150, 184 and 312 GHz when $f_{RF}$=0.918 GHz are shown in Fig. 5b-d, respectively. The 1 Hz linewidth of these spectra demonstrates the stable operation of the actively mode locked RTD-based THz frequency comb enabled by strong RF injection.

To verify the coherence of each comb mode under the active mode-locking condition, we conducted the phase noise measurement. As the comb modes are equally spaced by $f_{RF}$, their beat signals overlap at this frequency. Hence, the phase noise measured at $f_{RF}$ represents the superposition of noise contributions from all these interline beats, serving as an effective indicator of the overall coherence of the frequency comb. Based on this principle, we used an Schottky Barrier Diode (SBD) detector to receive comb signal and measured the phase noise at $f_{RF}$ using a phase noise analyzer to evaluate the coherence of the frequency comb. As shown in Fig. 5e, the measured phase noise at offset frequency of 1, 10 and 100 Hz was approximately -60, -80 and -100 dBc/Hz, respectively. The ultralow phase noise confirms the high coherence of individual comb modes and validates the successful generation of a frequency comb.

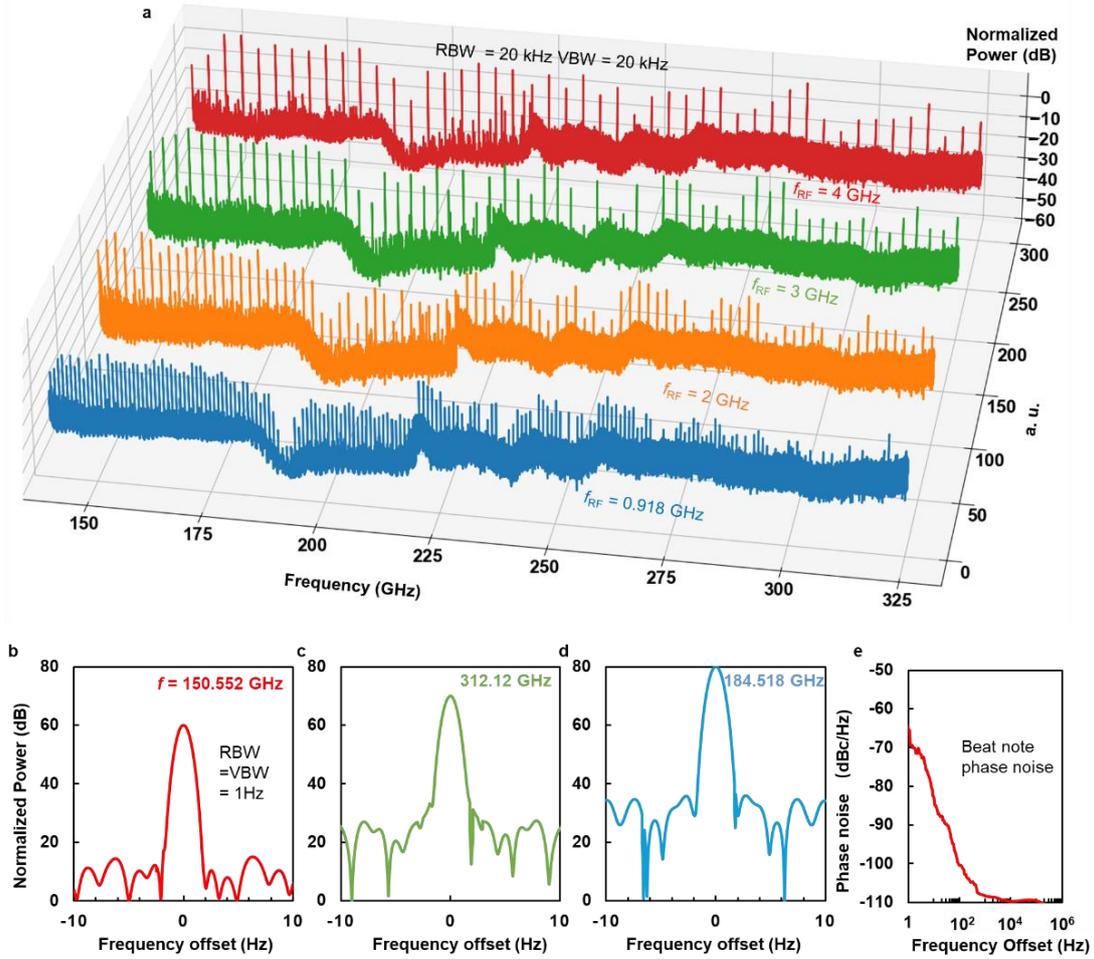

Fig.5 (a) Spectrum of frequency comb with RF injection power of 18 dBm and $f_{RF}$ = 0.918, 2, 3, 4 GHz respectively. Magnified spectra with $f_{RF}$ = 0.918 GHz at (b) 150, (c) 184 and (d) 312 GHz show 1 Hz linewidth respectively. (e) The measured phase noise of beat note generated by frequency comb.

As a proof of concept, we employed the THz frequency comb source as the transmitter in a THz transmission experiment. In the back-to-back configuration, the transmitter and receiver were directly connected via a waveguide. At the transmitter side, an RF signal at $f_{RF}$=20.6 GHz was used to generate three distinct carrier frequencies-$f_6$=123.6 GHz, $f_7$=144.2 GHz and $f_9$=185.4 GHz-for independent channel testing. The THz frequency comb was amplified by a 20-dB-gain THz low-noise amplifier before being fed into a fundamental mixer to upconvert a 5 GHz subcarrier 16-QAM baseband signal. Each comb mode carries identical baseband information. At the receiver side, the local oscillator (LO) signals at 123.6 GHz, 144.2 GHz, and 185.4 GHz were generated by multiplying and terahertz sub-harmonically mixing. The received THz signal was down-converted to 5 GHz via homodyne detection and subsequently amplified by a 38-dB-gain millimeter wave low-noise amplifier. The resulting 3 Gbaud 16-QAM baseband signals from the three comb-mode channels were captured using a real-time oscilloscope and are shown in the constellation diagrams of Fig. 6. Variations in error vector magnitude across the three channels are attributed to differences in the SNR among the respective comb modes. The achieved data rate of 12 Gbps per channel highlights the potential of this compact THz frequency comb source for high-speed, multi-channel THz transmission.

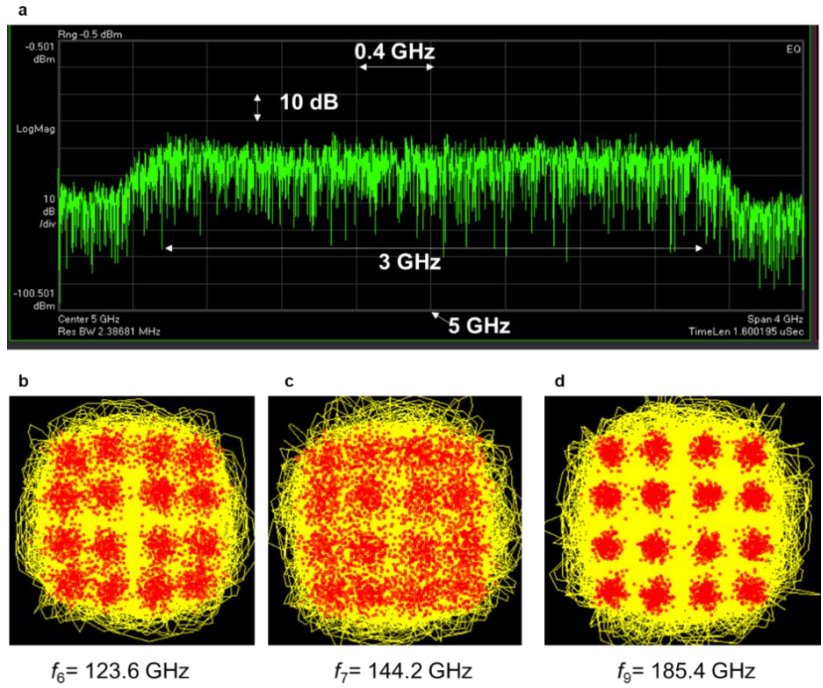

Fig.6 (a) Detected baseband signal with 3 Gbaud and 5 GHz subcarrier. (b) Constellation diagram of the demodulated 3 Gbaud 16-QAM signal using 123.6, 144.2 and 185.4 GHz channels, respectively.

## V. Conclusion

We demonstrated a first broadband THz frequency comb generated by a RTD under active mode-locking condition. Ultralow phase noise and sub-Hertz linewidth of each comb mode confirm the high coherence and validate the frequency comb operation. The comb spans a wide frequency range from 140 to 325 GHz, with a relatively flat spectral region from 140 to 190 GHz. The mode spacing is broadly tunable, ranging from 0.8 GHz to over 40 GHz. We utilize this comb source for multi-channel transmission experiments, achieving a data rate of 12 Gbps per channel. These results highlight the potential of RTD-based THz frequency comb as compact and tunable THz frequency comb sources, and future work will focus on enhancing the output power of frequency comb for THz applications. These results indicate that RTD-based THz frequency comb is a very promising candidate for realizing compact concepts in various integrated THz application systems.